\documentclass[prl,superscriptaddress,twocolumn]{revtex4-1}
\usepackage{graphicx}
\usepackage{bm}
\usepackage{amsmath}
\usepackage{amssymb}
\usepackage{exscale}
\usepackage{wasysym}
\usepackage{ulem}
\usepackage{cancel}
\bibstyle{apsrev.bib}

\usepackage{xcolor}

\begin{document}


\title{Shear flow of non-Brownian rod-sphere mixtures near jamming}

\author{Carmine Anzivino}

\email[Electronic mail: \ ]{carmine.anzivino@unimi.it,}

\affiliation{Department of Physics ``A. Pontremoli", University of Milan, via Celoria 16, 20133 Milan, Italy}

\author{Christopher Ness}
\email[Electronic mail: \ ]{chris.ness@ed.ac.uk,}

\affiliation{School of Engineering, University of Edinburgh, Edinburgh, EH9 3JL, United Kingdom}

\author{Amgad Salah Moussa}

\affiliation{Syngenta AG, 4058 Basel, Switzerland}

\author{Alessio Zaccone}

\email[Electronic mail: \ ]{alessio.zaccone@unimi.it}

\affiliation{Department of Physics ``A. Pontremoli", University of Milan, via Celoria 16, 20133 Milan, Italy}

\begin{abstract}
We use the discrete element method, taking particle contact and hydrodynamic lubrication into account, to unveil the shear rheology of suspensions of frictionless non-Brownian rods in the dense packing fraction regime. We find that, analogously to the random close packing volume fraction, the shear-driven jamming point of this system varies in a non-monotonic fashion as a function of the rod aspect ratio. The latter strongly influences how the addition of rod-like particles affects the rheological response of a suspension of frictionless non-Brownian spheres to an external shear flow. At fixed values of the total (rods plus spheres) packing fraction, the viscosity of the suspension is reduced by the addition of ``short" ( $\leq 2$) rods but is instead increased by the addition of ``long" ( $\geq2$) rods. A mechanistic interpretation is provided in terms of packing and excluded-volume arguments.
\end{abstract}

\maketitle


Suspensions of non-Brownian, micron-sized, particles dispersed in Newtonian fluid are ubiquitous in nature and have widespread industrial applications, especially in the \textit{dense} regime where solid and fluid are mixed in similar proportions \citep{guazzelli_pouliquen_2018,Chris_review,Wu1,Wu2}. 
The application of a shear deformation leads to a (shear-driven) \textit{jamming} transition upon increasing the solid packing fraction $\phi$ towards a so-called jamming point $\phi_\textrm{J}$ \citep{YuliangJIN_review, Olsson_Teitel}. While a dense non-Brownian suspension can flow under an external shear stress for $\phi < \phi_\textrm{J}$, the viscosity $\eta$ of the suspension increases dramatically when $\phi \to \phi_\textrm{J},$ and the system consequently develops a solid-like behavior with a finite yield stress at $\phi \ge \phi_\textrm{J}.$ Although the mechanism underlying this flow arrest is not yet understood, the jamming transition is commonly believed to crucially influence the shear rheology of non-Brownian suspensions in the dense regime (i. e. below but not too far from $\phi_\textrm{J}$) of packing fraction \citep{Heussinger_Barrat, Andreotti_2012, Chris_review}.

When the particles are frictionless and spherical, the viscosity $\eta$ of the suspension exhibits a power law divergence $\eta (\phi) \approx  (\phi_\textrm{J} - \phi)^{-\beta},$ where $\beta$ is a scaling parameter much discussed in the literature 	\citep{Andreotti_2012} and reported to be  $\beta \approx 2$ in shear flow experiments \citep{Guy_Hermes_Poon}, while $\phi_\textrm{J}$ coincides with the \textit{random close packing} (RCP) volume fraction, $\phi_\textrm{RCP} \approx 0.64,$ of a collection of hard spheres \citep{Olsson_Teitel}.  
The value of $\phi_\textrm{RCP}$ for a generic ensemble of hard particles is defined as the highest packing fraction for a ``disordered" arrangement of those particles~\citep{Anzivino}. Since non-sheared systems in the liquid phase for $\phi < \phi_\textrm{RCP}$ reach mechanical rigidity at $\phi_\textrm{RCP}$, the latter quantity is often referred to as the jamming point~\citep{Hern_jamming,Liu_Nagel_2010}. Nevertheless, by contrast with $\phi_\textrm{J},$ $\phi_\textrm{RCP}$ for a given system is measured by means of an isotropic compression rather than a shear deformation. To highlight this difference, some authors have recently referred to the transitions occurring at $\phi_\textrm{J}$ and $\phi_\textrm{RCP}$ as \textit{shear-driven jamming} and  \textit{compression-driven jamming}, respectively \citep{teitel_2019,teitel_2020}. 


\begin{figure}
\centering
\includegraphics[width = 0.8 \linewidth]{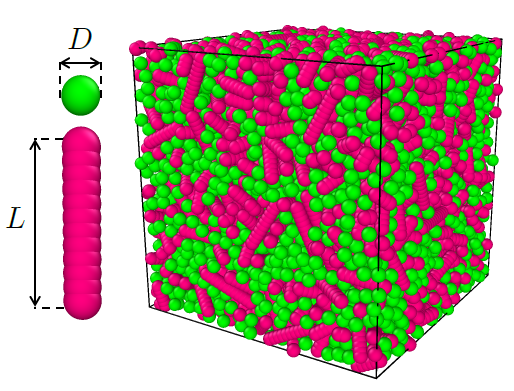}
	\caption{Randomly packed rod-sphere mixture. The spheres have diameter $D,$ while the rods (spherocylinders formed by glued spheres) have aspect ratio $L/D=4.$ Packing fraction is $\phi=0.6.$}
		\label{sketch}
\end{figure}

\begin{figure*}[ht]
\centering
\includegraphics[width = 1 \linewidth]{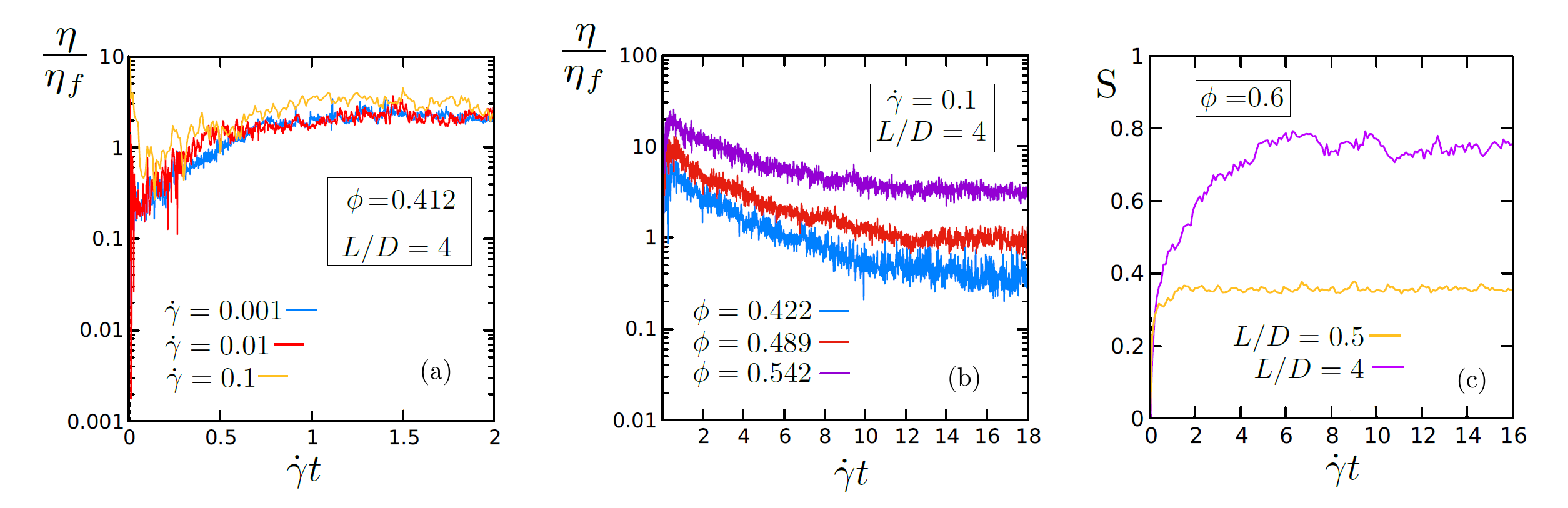}
	\caption{$(a)$ Dimensionless viscosity $\eta / \eta_\textrm{f}$ of a suspension of pure (no spheres in the mixture) spherocylinders, fixed aspect ratio $L/D=4$ and packing fraction $\phi=0.412,$ as a function of the shear strain $\dot{\gamma} t$ for several values of the shear rate $\dot{\gamma}.$ $(b)$ After a start up regime in which the viscosity increases with strain, a plateau is reached. At this plateau, orientational order (negligible for short ARs) has been developed as proved by a plot of the nematic scalar order parameter $S$ as a function of $\dot{\gamma} t$, $(c).$ The shear rate is measured in simulation units.}
 		\label{Fig1}
\end{figure*}


An increasing number of numerical and experimental studies have recently investigated how $\phi_\textrm{RCP}$ of Brownian and non-Brownian suspensions is influenced by the nonspherical shape of the dispersed particles \citep{Philipse1,PhilipsePSS,donev,Sacanna_2007,Sacanna_RodSphere}. How this latter property affects the shear-driven jamming point $\phi_\textrm{J}$ and the related shear rheology in the region $\phi < \phi_\textrm{J}$ has remained, instead, poorly understood. Even less attention has been devoted to exploring how the shear rheology varies when particles with different shapes are dispersed within the same suspension.


To fill this gap, we consider a mixture of spheres and rods under simple shear flow. Rods are modelled as spherocylinders: axially symmetric cylinders of length $L$ and diameter $D$, capped by hemispheres also of diameter $D$ (see Fig. \ref{sketch}). We compute how the viscosity $\eta$ of the mixture varies as a function of both the aspect ratio (AR) $L/D$ of the rods and the relative concentration $x$ of the spheres. We employ a recently introduced numerical method \citep{Cheal}, which is based on an analogy between dense non-Brownian suspensions and dry granular matter. This analogy exploits the fact that, by contrast with dilute regimes where long-range hydrodynamic interactions, particle contact interactions and random packing dominate the dynamics in the dense regime near jamming \citep{Chris_review}.

\begin{figure*}[ht]
\centering
\includegraphics[width = 1.0 \linewidth]{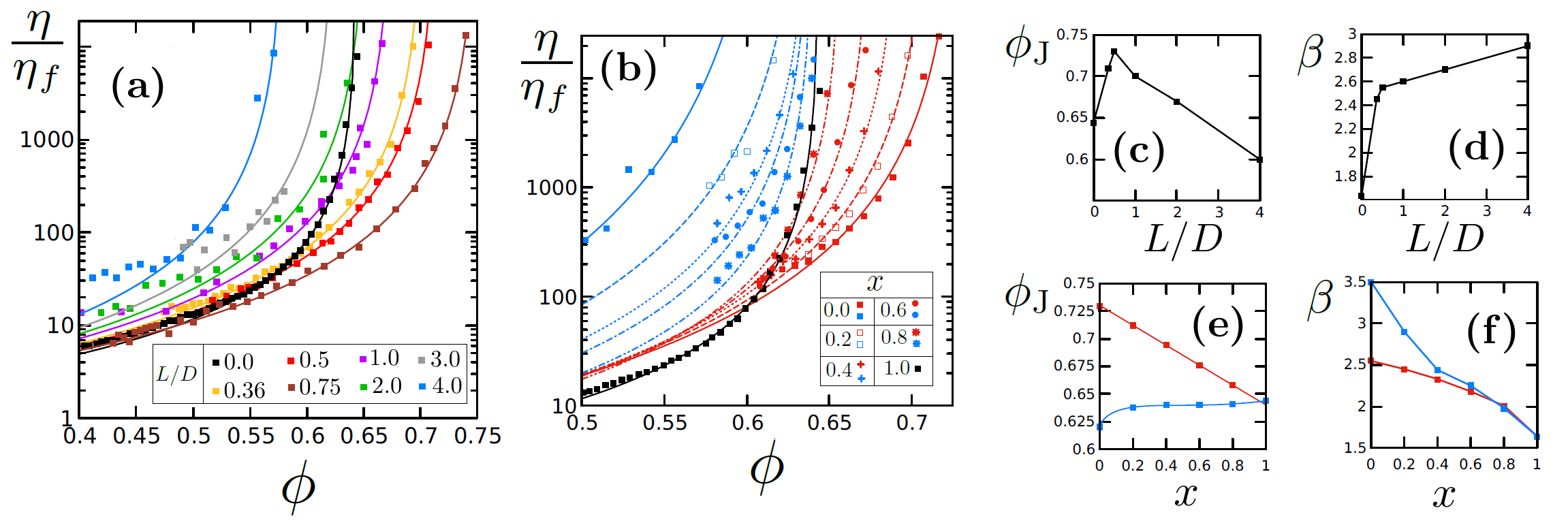}
	\caption{Shear viscosity $\eta/\eta_f$ of a rod-sphere mixture as a function of the total (rods plus spheres) particle concentration $\phi$, for several values of the rod aspect ratio $L/D$ at fixed concentration of spheres $x=0$ $(a),$ and for several values of $x$ at fixed $L/D=0.5$ (red points) and $L/D=4$ (blue points) $(b).$ All data can be fitted to the KD relation (Eq. \eqref{KD}). The $\phi_\textrm{J}$ and $\beta$ coefficients of the curves in Fig. $(a)$ are plotted in $(c)$ and $(d),$ respectively. The $\phi_\textrm{J}$ and $\beta$ coefficients of the curves in Fig. $(b),$ i. e. at fixed $L/D=0.5$ (red) and $L/D=4$ (blue), are plotted in $(e)$ and $(f),$ respectively.}
		\label{Fig2}
\end{figure*}




Unlike spheres, rods behave differently when subjected to isotropic compression or to shear. While in the former case they show no long-range orientational order upon jamming, in the latter case they do show orientational ordering as a result of torques induced by the shear flow. This shear-induced orientational ordering has been widely demonstrated both in simulations \citep{stannarius_NEMATIC,stannarius_NEMATIC_PRE,Guo} and in experiments \citep{stannarius_NEMATIC, stannarius_NEMATIC_PRE, Sandra}. Comparatively, little attention has been devoted to understanding the implications of this ordering on the shear-induced jamming point $\phi_\textrm{J}.$ In simulations of bidisperse mixtures of two-dimensional spherocylinders, Refs. \citep{teitel_2019,teitel_2020} have found a monotonic increase of $\phi_\textrm{J}$ with the AR.

Here we show that, analogously to the random close packing volume fraction $\phi_\textrm{RCP}$, the shear-driven jamming point $\phi_\textrm{J}$ of a suspension of spherocylinders also varies non-monotonically as a function of the AR $L/D$. Furthermore, we show that the addition of rods strongly affects the rheological response of a suspension of frictionless non-Brownian spheres to an external shear stress.

Our simulations employ a well-established code implemented in LAMMPS \citep{Ness2023}.
We simulate the trajectories of mixtures of $N_{S}$ spheres and $N_{R}$ rods in a periodic cubic box of side $L^B.$
Rods (spherocylinders) consist of linear assemblies of spheres (diameter: $D$, density: $\rho$, and  stiffness: $k_n$) of varying length $L$. To ensure bulk conditions, we ensure $L^B \gg D$ and $L^B \gg L.$ 
Spheres follow Newtonian dynamics, subject to forces and torques arising due to Stokes drag, hydrodynamic lubrication and repulsive contact.
Drag forces are computed relative to a background streaming fluid flow (viscosity $\eta_f$) so that a linear velocity profile $\mathbf{u}^{\infty} = (\dot{\gamma} y, 0, 0)$ is established with shear rate $\dot{\gamma}$.
The lubrication forces are computed according to Ref. \citep{Jeffrey_Onishi_1984}, and they are truncated at $0.001 (D/2)$ to prevent divergence.
Contacts are modelled as stiff linear springs with repulsive force set by the sphere-sphere overlap and stiffness. We indicate the stress tensor by $\Sigma.$
Full details of the forces and torques are given by~\citet{Cheal}.

For spheres, we sum the force at each timestep and update the acceleration according to the Velocity-Verlet algorithm.
For rods, we sum the forces over all constituent spheres then distribute the resultant force uniformly to each sphere.
This ensures that rods act as rigid bodies with no relative translation or rotation between constituent spheres.
Further details are given in Ref.~\citep{Blair_Ness_2022}.
We set $\rho\dot{\gamma}(D/2)^2/\eta_f <10^{-2}$ and $\dot{\gamma}\sqrt{\rho (D/2)^3/k_n}<10^{-4}$ to ensure, respectively, inertia-free and hard sphere conditions. The total packing fraction is $\phi = \phi_S + \phi_R,$ where $\phi_S = (4/3) \pi (D/2)^3 \rho$ and $\phi_R = (4/3) \pi (D/2)^3 \rho_R + \pi (D/2)^2 \rho L.$
Moreover $x \equiv \phi_S / \phi$ is the relative component of spheres in the mixture, and $\phi_R / \phi = 1 -x$.
To simulate simple shear we use a triclinic periodic box with a tilt length $L_{xy}^B$ that is incrementally increased linearly in time as $L_{xy}^B (t) = L_{xy}^B (t_0) + L_{y}^B \dot{\gamma} t,$ giving a deformation equivalent to that obtained using Lees-Edwards conditions. The viscosity of the mixture is $\eta = \Sigma_{xy} / ( \dot{\gamma} \eta_f ),$ where $\Sigma_{xy}$ is the $xy$ component of $\Sigma$ and $\eta_\textrm{f}$ is the viscosity of the hosting fluid.
We thus compute the viscosity $\eta$ of mixtures of spheres and rods, for several combinations of the total (spheres plus rods) particle packing fraction $\phi \equiv \phi_S+ \phi_R,$ the relative concentration of the spheres $x \ (\equiv \phi_S / \phi),$ and the AR of the rods $L/D.$ 

We start from the case $x=0,$ corresponding to a monodisperse ensemble of rods (for which $\phi_S=0$ and $\phi \equiv \phi_R,$ respectively). In Fig. \ref{Fig1}$(a),$ we plot the dimensionless viscosity $\eta / \eta_\textrm{f}$ at fixed $\phi=0.412$ and $L/D=4$ as a function of the shear strain $\dot{\gamma} t,$ for several values of the shear rate $\dot{\gamma}.$ As it can be observed, our results do not depend on the chosen value of $\dot{\gamma}.$ Fig. \ref{Fig1}$(b)$ shows that after a start up transient in which the viscosity increases with strain, a plateau is reached. At this plateau, orientational order (negligible for short ARs) has developed in the system. This is shown in Fig. \ref{Fig1}$(c)$ where the nematic scalar order parameter $S$ is plotted as a function of $\dot{\gamma} t.$
To compute the shear-induced orientational order we diagonalize the  tensor 
$Q_{\alpha \beta} = \frac{1}{N_R} \sum_{i=1}^{N_R} \bigg( \frac{3}{2} u_\alpha^i u_\beta^i - \frac{\delta_{\alpha \beta}}{2} \bigg)$,
where $\alpha, \beta=x,y,z.$ In this definition, $\mathbf{u}^i$ is a unit vector along the long axis of particle $i,$ and the sum is considered over all the $N_R$ rods in the mixture. The scalar nematic order parameter $S$ is defined as the largest eigenvalue of $\mathbf{Q}.$

In all cases the viscosity is measured at a value of strain $\dot{\gamma} t$ sufficiently large for the plateau of Fig. \ref{Fig1}$(b)$ to be reached. For several values of $L/D \in [0, 4],$ we plot the viscosity $\eta$ of this system as a function of $\phi,$ with symbols in Fig. \ref{Fig2}$(a).$ In all cases, we express $\eta$ rescaled by $\eta_f.$ In the figure, black points represent the limiting case $L/D=0,$ which corresponds to a monodisperse suspension of non-Brownian frictionless spheres. The black points in Fig. \ref{Fig2}$(a)$ can be fitted to the widely used Krieger-Dougherty (KD) relation
\begin{equation} \label{KD}
\eta / \eta_f = \alpha (1- \phi / \phi_\textrm{J})^{-\beta},
\end{equation}
with fitting parameters $\alpha=1,$ $\beta=1.6$ and $\phi_\textrm{J} = 0.644,$ respectively. 
Moreover we show that, besides the case $L/D=0,$ all the numerical data plotted in Fig. \ref{Fig2}$(a)$ can be fitted to the KD relation \eqref{KD}, once the values of $\beta$ and $\phi_\textrm{J}$ are properly fitted. All the fitted curves are plotted with full lines in Fig. \ref{Fig2}$(a)$ and Fig. \ref{Fig2}$(b)$. The values of $\beta$ and $\phi_\textrm{J}$ are plotted as a function of $L/D$ in Figs. \ref{Fig2}$(c)$ and \ref{Fig2}$(d),$ respectively. Analogously to $\phi_\textrm{RCP},$ we find $\phi_\textrm{J}$ to vary non-monotonically as a function of $L/D,$ with a maximum reached at $L/D=0.75.$

\begin{figure}[ht]
\centering
\includegraphics[width = 0.85 \linewidth]{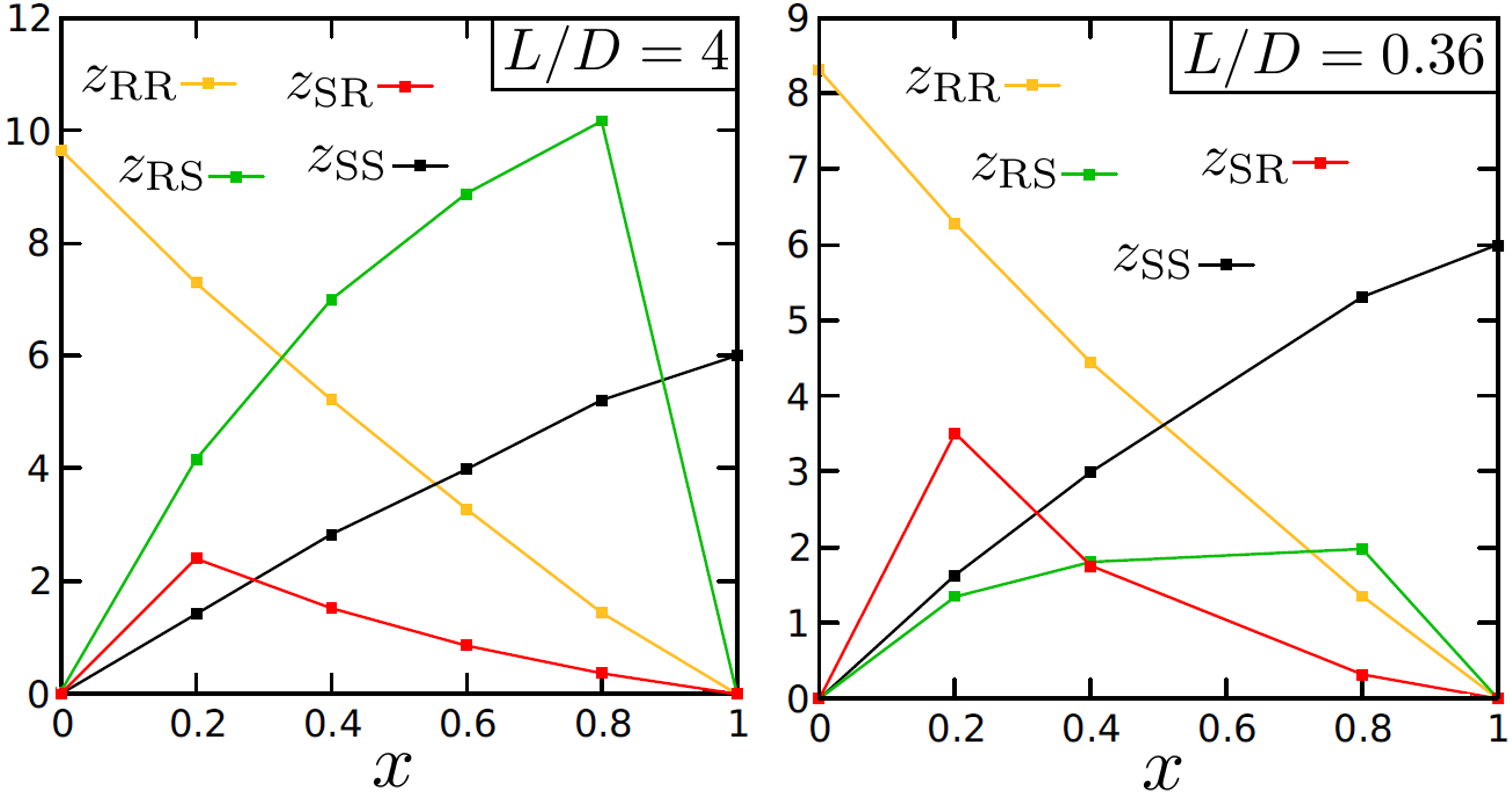}
	\caption{Mean contact numbers ($z$) in a rod-sphere mixture at the shear-driven jamming point $\phi_\textrm{J},$ versus the concentration $x$ of spheres in the mixture. The number of rods in contact with each rod is $z_\textrm{RR},$ the number of rods in contact with each sphere is $z_\textrm{RS},$ the number of spheres in contact with each rod is $z_\textrm{SR}$ and the number of spheres in contact with each sphere is $z_\textrm{SS}.$ Left: the rods have aspect ratio $L/D=4$. Right: the rods have aspect ratio $L/D=0.36.$}
		\label{contact}
\end{figure}


After having explored the case $x=0,$ corresponding (for $L/D >0$) to the absence of spheres in the system, we consider a more complex situation in which both rods and spheres are present in the sample so that $x \neq 0.$ As for the case of monodisperse rods, for rod-sphere mixtures it is convenient to use a collection of frictionless spheres as a reference state. The latter situation can be obtained in two distinct ways: either (i) choosing the value of spheres concentration $x=1$ independently of the AR of the rods, or (ii) fixing $L/D=0$ independently of the spheres concentration $x.$ In both cases, the dependence of the viscosity $\eta$ on $\phi$ is given by the black curve depicted in Fig. \ref{Fig2}$(a).$
To explore other non-trivial situations, while keeping the AR $L/D$ of the rods fixed, we compute the viscosity $\eta$ of the mixture as a function of $\phi,$  for several values of $x \in [0,1].$ We repeat this protocol for (almost) the same values of AR considered in Fig. \ref{Fig2}$(a),$ for $L/D \in ]1,5].$ The results can be already guessed from Fig. \ref{Fig2}$(a).$ Each curve plotted in this figure, indeed, represents the case $x=0$ at fixed $L/D.$ The black curve, instead, represents the case $x=1$ for any value of $L/D.$ It follows that all the ``colored" curves must collapse on the black one by increasing the relative concentration $x$ of the spheres in the mixture. 

Thus, by adding rods to spheres, two contrasting behaviors are observed, depending on $L/D.$ While addition of short rods ($L/D < 2$) results in a decrease of the viscosity at fixed $\phi,$ addition of long rods ($L/D \ge 2$) results in an increase. This scenario is depicted in Fig. \ref{Fig2}$(b),$ illustratively for the case of rods with AR $L/D=0.5$ (red points) and $L/D=4$ (blue points), respectively. Full lines represent fits to the KD relation \eqref{KD} whose parameters are reported in Figs. \ref{Fig2}$(e)$ and \ref{Fig2}$(f),$ respectively.

This result can be interpreted in terms of the average number of contacts per particle, $z$, required for a rod-sphere mixture to be mechanically stable. The average contact numbers for the various species computed from the simulations are shown in Fig. 4, as a function of the concentration of spheres $x$, for two ARs.

Again it is instructive to start from pure rods, $x=0.$ As it is known \citep{donev,Hoy_2023}, the average critical contact number $z_\textrm{J}$ in a monodisperse system of jammed rods increases from the value $z=6$ at $L/D=0$ until a value $z_\textrm{J} \approx 10$ is reached at approximately $L/D =0.5$ and then remains constant upon further increment of $L/D.$ As argued in \cite{donev}, the way $z_\textrm{J}$ varies as a function of $L/D$ provides an explanation for the non-monotonic variation (shown in Fig. \ref{Fig2}$(c)$) of $\phi_\textrm{J}$ on $L/D$, which is confirmed here, for the first time, for jamming under shear flow.  More specifically, the number of degrees of freedom per particle in the system increases with $L/D$ as rotational degrees of freedom add to the translational ones when rods replace spheres \cite{Zaccone_book}. The increase in the number of degrees of freedom per particle results in an increase of the overall number of particle contacts $z_\textrm{J}$ required to mechanically stabilize the packing as $L/D$ increases. In turn, an increase of $z_\textrm{J}$, at the onset of mechanical stability is associated with an increase of the corresponding $\phi_\textrm{J}$ \cite{Zaccone_PRL_RCP,Hoy_2023}. After a certain threshold ($L/D \approx 0.5$) is reached, the number of degrees of freedom does not depend on $L/D$ anymore, then $z_\textrm{J}$ remains constant after the value $z=10$ has been reached. In this situation, the subsequent decrease of $\phi_\textrm{J}$ for markedly aspherical particles in orientationally disordered packings is explained by strong excluded-volume effects \`a la Onsager \cite{Onsager,Hoy_2023}. This argument explains the non-monotonic trend of $\phi_\textrm{J}$ of rod-sphere mixtures vs $L/D$ also in the presence of shear flow, observed here for the first time.

A quantitative explanation of the dependence of the viscosity $\eta$ of the system on $L/D$ follows from $\phi_\textrm{J}$ being the point at which $\eta$ diverges (from Eq. \eqref{KD} and as numerically shown in the above). In particular, in the first regime where $\phi_\textrm{J}$ increases with increasing $L/D$, the viscosity decreases because the viscosity is always lower when $\phi_\textrm{J}$ is larger, cfr. also Eq. \eqref{KD}. Conversely, when $\phi_\textrm{J}$ decreases with increasing $L/D$ (excluded-volume effects \`a la Onsager), for the reasons explained above, then the viscosity increases with further increasing $L/D$. However, rather than at $L/D=0.5,$ we find the maximum value of $\phi_\textrm{J}$ to be located at $L/D=0.75.$

To summarize, for a system of pure rods there are two different mechanisms which determine the location of the jamming point $\phi_\textrm{J}$ as a function of the AR $L/D.$ For short rods excluded-volume effects \`a la Onsager are not dominant, the increase of $\phi_\textrm{J}$ as a function of $L/D$ is caused by the increase of $z$ due to the emergence of additional rotational degrees of freedom. Excluded-volume effects \`a la Onsager instead dominate in orientationally disordered packings of long rods \cite{Hoy_2023}, from which a decrease of $\phi_\textrm{J}$ in the region $L/D \ge 0.75$ arises. The trend of $\phi_\textrm{J}$ then determines the trend of the viscosity versus $L/D$, which will be opposite, i.e. anti-correlated, to that of $\phi_\textrm{J}$, according to Eq. \eqref{KD}.

This scenario can be generalized to the case of a rod-sphere mixture. In this case, four contact numbers exist, see Fig. \ref{contact}. 
When spheres are added to a system of randomly jammed \emph{long} rods, they mainly act as to fill the large voids between the rods. As a consequence, the jamming point $\phi_\textrm{J}$, at which the viscosity diverges, increases upon increasing the fraction of spheres, and the viscosity decreases. By contrast, when spheres are added to a system of randomly jammed \emph{short} rods, they mainly act as to reduce the number of degrees of freedoms in the system, by effectively ``killing'' the rotational degrees of freedom. As a consequence, the jamming point $\phi_\textrm{J}$ decreases with the addition of spheres, in this regime, and the viscosity increases. 

In summary, we unveiled the shear rheology of a binary mixture of spheres and rods (spherocylinders) numerically. Our main finding is that the effect of adding rods on the viscosity of the mixture strongly depends, in a non-monotonic fashion, on the rod aspect ratio AR. Adding rods to spheres reduces the viscosity of the suspension as long as the rods have AR $L/D<2,$ with a minimum value of the viscosity at $L/D=0.75.$ When rods with $L/D \ge 2$ are added, instead, the viscosity dramatically increases.
Our findings pave the way for the rational control of viscosity for energy-saving purposes. 

\subsection{Acknowledgements}
A.Z. gratefully acknowledges funding from the European Union through Horizon Europe ERC Grant number: 101043968 ``Multimech'', and from US Army Research Office through contract nr. W911NF-22-2-0256. C.A. gratefully acknwoledges funding from Syngenta AG (Switzerland) and Syngenta U.K.

\bibliographystyle{apsrev4-1}

\bibliography{references}

\end{document}